\pdfoutput=1
\documentclass[11pt,reqno,preprint,dvipsnames]{article}
\usepackage{jheplike,epsfig,amssymb,amsmath,mathrsfs,hyperref,array,hhline}
\usepackage[makeroom]{cancel}
\usepackage{multirow,bbm}
\usepackage{feynmp-auto}
\usepackage{cleveref}
\usepackage[textsize=tiny]{todonotes}
\usepackage{mathtools}
\usepackage{caption}
\usepackage{subcaption}
\usepackage{placeins}
\usepackage{marginnote}

\usepackage{tikz}
\usetikzlibrary{tikzmark}
\usetikzlibrary{decorations.markings}
\usetikzlibrary{shapes}
\usetikzlibrary{arrows.meta}

\usetikzlibrary{decorations.pathreplacing,calc,arrows,decorations.markings,decorations.pathmorphing}

\newcommand{\lefttodo}[2][]{{%
 \let\marginpar\marginnote
 \reversemarginpar
 \renewcommand{\baselinestretch}{0.8}%
 \todo[#1]{#2}}}

\makeatletter
\DeclareRobustCommand*{\bfseries}{%
  \not@math@alphabet\bfseries\mathbf
  \fontseries\bfdefault\selectfont
  \boldmath
}

\def\be{\begin{equation}}
\def\ee{\end{equation}}
\newcommand{\bea}{\begin{eqnarray}}
\newcommand{\eea}{\end{eqnarray}}

\def\Hhex{{\cal H}^{\rm hex}}

\DeclareMathOperator{\Div}{Div}
\DeclareMathOperator{\Fin}{Fin}

\title{Solving Scattering in $\mathcal{N}=4$ Super-Yang-Mills Theory}

\author{Nima Arkani-Hamed$^{1}$, Lance J.~Dixon$^{2}$, Andrew J.~McLeod$^{3,4}$, Marcus Spradlin$^{5,6}$, Jaroslav Trnka$^{7}$, Anastasia Volovich$^{5}$}

\affiliation{$^1$ School of Natural Sciences, Institute for Advanced Study,
Princeton, NJ 08540, and
Center of Mathematical Sciences and Applications, Harvard University,
Cambridge, MA 02138}

\affiliation{$^{2}$ SLAC National Accelerator Laboratory,
Stanford University, Stanford, CA 94309, USA}

\affiliation{$^{3}$ CERN, Theoretical Physics Department, 1211 Geneva 23, Switzerland}

\affiliation{$^{4}$ Mani L. Bhaumik Institute for Theoretical Physics, \\ UCLA Department of Physics and Astronomy, Los Angeles, CA 90095, USA}

\affiliation{$^{5}$ Department of Physics, Brown University, Providence, RI 02912, USA}

\affiliation{$^{6}$ Brown Theoretical Physics Center, Brown University, Providence, RI 02912, USA}

\affiliation{$^{7}$ Center for Quantum Mathematics and Physics (QMAP),
Department of Physics, University of California, Davis, CA 95616, USA}

\abstract{
As part of the Snowmass community planning exercise, we highlight an
ongoing program of research into the structure of scattering amplitudes
in ${\cal N}=4$ super-Yang-Mills theory, particularly in the planar limit
of a large number of colors.
This theory sits at the nexus of a number of exciting topics in
high-energy particle physics, including the AdS/CFT correspondence,
conformal field theory, integrability, and string theory, and is believed
to be exactly solvable in four dimensions.  In many ways,
planar ${\cal N}=4$ super-Yang-Mills theory is the ``hydrogen atom''
of relativistic scattering:  It has proven indispensable
for learning about new geometrical formulations of quantum field theory,
for exploring mathematical properties at high perturbative orders,
and for developing powerful new computational methods that have found
applicability in precision collider physics.}

\preprint{\vspace*{-2cm}  \begin{flushright} CERN-TH-2022-123 \\ SLAC-PUB-17692  
 \end{flushright}}

\begin{document}
\maketitle
\flushbottom
\begin{fmffile}{feyndiags}


\section{Introduction}
\label{sec:introduction}

The study of scattering amplitudes has played a central role in the
development of theoretical physics, and has led to some of the most precise predictions in all of
science~\cite{Aoyama:2012wj,Aoyama:2017uqe,Laporta:2017okg}. 
These predictions have traditionally been made with the use of Feynman diagrams, which provide 
an intuitive picture for scattering amplitudes as the sum over all ways a given configuration of incoming particles can scatter into a configuration of outgoing particles. However, we now know
that there are completely different ways of formulating scattering
amplitudes that make no reference to particle trajectories, or even any
notion of space-time. These novel ways of thinking about scattering amplitudes have mainly arisen from investigations of
${\cal N}=4$ supersymmetric Yang-Mills (SYM) theory in four dimensions, the theory that we focus on in this white paper.

Part of the motivation for recasting scattering amplitudes in new and more abstract ways comes from the incredible simplicity these quantities exhibit,
relative to the complexity of the calculations currently required to compute them.  
This has been especially true in ${\cal N} = 4$ SYM theory, in which seemingly-miraculous cancellations have 
led to the discovery of beautiful mathematical structures that make contact with many branches of modern mathematics, 
including combinatorics, algebraic geometry, number theory, and the theory of motives. The endeavor to understand the 
simplicity of amplitudes has correspondingly led to a rich and productive interplay between amplitudes researchers and mathematicians. 

\begin{figure}[t]
\begin{center}
\includegraphics[width=6.5in]{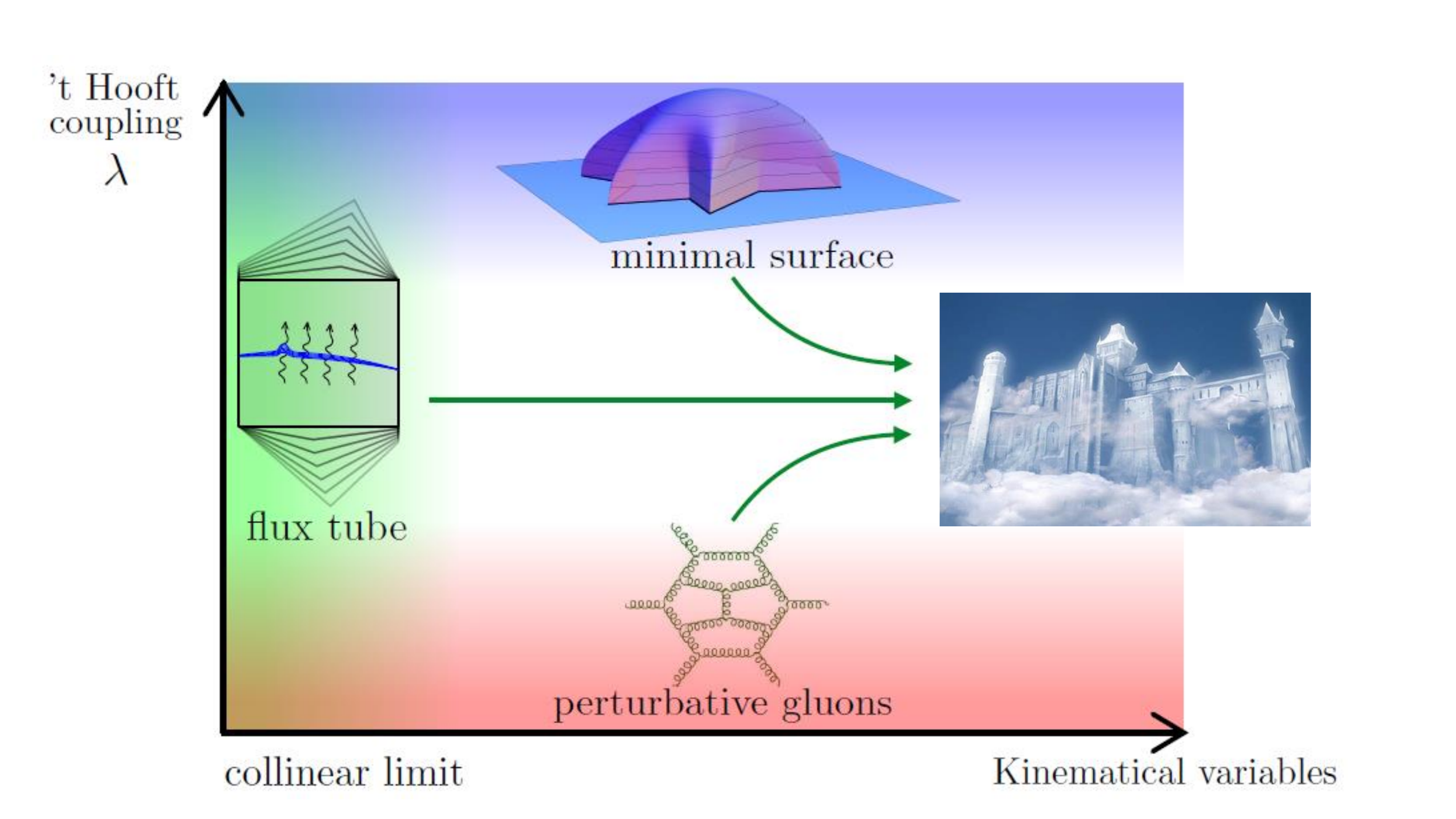} 
\end{center}
\caption{The three approaches that should be unified to solve
${\cal N}=4$ super-Yang-Mills theory in the planar limit:
weak coupling via perturbation theory, strong coupling via minimal surfaces,
and near-collinear kinematics via the pentagon operator product expansion
at any coupling.}
\label{Fig:SolveNeq4Figure}
\end{figure}

Our understanding of the {\it planar} limit of ${\cal N}=4$ SYM theory, in which the number of colors in the SU($N_c$) gauge group becomes large, 
is especially well developed. There are currently three independent descriptions of scattering
amplitudes in this regime, illustrated in Figure~\ref{Fig:SolveNeq4Figure}.
A weak-coupling formulation makes
contact with perturbative methods involving Feynman diagrams~\cite{Feynman:1949zx,Caron-Huot:2020bkp};
a ``holographic" strong-coupling formulation employs minimal-area surfaces
in Anti-de Sitter space~\cite{Alday:2007hr,Alday:2010vh};
and a pentagon operator product expansion (POPE)
approach exploits the two-dimensional integrability of a dual
string picture at finite coupling in various kinematic
limits~\cite{Alday:2010ku,Basso:2013vsa,Basso:2014pla}.
These formulations are all mutually consistent but make use of different physics, 
are formulated in mathematically distinct ways, and make different properties of amplitudes manifest.
A major task for the future will be to find a single unifying description of amplitudes in this theory that 
properly matches each of these formulations in the appropriate limit. Mathematically, this question 
can be framed as a search for the functions that are able to express the markedly varied behavior exhibited by amplitudes,
from weak to strong coupling and in arbitrary kinematics.

Many interesting facets of these questions deserve attention in the coming
years. They include:
How do gluonic and stringy descriptions morph into each other
as the coupling and kinematics are varied?  What kind of
singularities show up, and what is the physics associated to them? 
How do holographic dualities, string theory, and even space-time itself,
emerge dynamically from planar gauge theories?
Solving scattering in planar ${\cal N}{=}4$ SYM theory will provide a
quantitative test for our physical and mathematical expectations, and will lead to an improved
intuition that can be applied to more general and realistic quantum field theories.

More immediately, the goal of this white paper is to lay out a set of concrete goals that are within the reach of 
current technology, and that will allow us to make progress on these overarching questions. Before outlining these goals, we provide a brief overview of the state of knowledge about ${\cal N}{=}4$ SYM theory, starting with a review of its particle content and symmetries in section~\ref{sec:background}. In section~\ref{sec:amplituheron} we describe what is known about amplitudes in this theory at the level of the integrand, while their properties as functions are described in 
section~\ref{sec:mathematical_properties}. The more detailed understanding we have of certain kinematic limits is described in section~\ref{sec:kinematic_limits}, and in section~\ref{sec:bootstrap} we explain how this understanding can be combined with 
knowledge of the analytic properties of amplitudes to in some cases bootstrap them directly. Finally, we highlight some of the research questions that we expect will be important in the coming decade in section~\ref{sec:outlook}.


\section{${\cal N} = 4$ Supersymmetric Yang-Mills Theory}
\label{sec:background}

The field content of SYM theory in four dimensions consists of a gauge field $A_\mu$, four Weyl fermion gluinos $\psi^{a A}$, and six scalar fields that are conventionally packaged into a two-index antisymmetric field $\phi^{AB} = - \phi^{BA}$.  The indices $A,B$ are vector indices of an unbroken $SU(4)$ R-symmetry that is possessed by the theory.  Since supersymmetry transformations relate the fields to each other, gauge invariance requires that $\psi^{a A}$ and $\phi^{AB}$ must transform in the adjoint representation of the gauge group, similar to $A_\mu$.  The $\mathcal{N}=4$ supersymmetric (indeed, superconformal) Lagrangian~\cite{Brink:1976bc} for this field content is unique up to the choice of gauge group and the value of a single complex, dimensionless coupling constant.

The on-shell degrees of freedom of the $\mathcal{N} = 4$ supermultiplet consist of a positive helicity gluon $g^+$, four $+\frac{1}{2}$ helicity gluino states $\tilde{g}_A$, six scalars $S_{AB}$, four $-\frac{1}{2}$ helicity gluino states $\smash{\overline{\tilde{g}}^A}$, and the negative helicity gluon $g^-$.  It is useful to package this collection of on-shell states into an on-shell superfield~\cite{Nair:1988bq}
\begin{equation}
\Phi(p^{a \dot{a}}, \eta^A) = g^+(p) + \eta^A \tilde{g}_A(p)
+ \tfrac{1}{2} \eta^A \eta^B S_{AB}(p)
+ \tfrac{1}{6} \eta^A \eta^B \eta^C \epsilon_{ABCD} \overline{\tilde{g}}^D(p)
+ \tfrac{1}{24} \eta^A \eta^B \eta^C \eta^D \epsilon_{ABCD} g^-(p)\,,
\end{equation}
in terms of which superamplitudes are constructed. This object is a function of an on-shell four-momentum $p$, which is often parametrize in terms of spinor helicity variables as
\begin{equation}
\label{eq:spinorhelicity}
p^{a \dot{a}} = \lambda^a \widetilde{\lambda}^{\dot{a}}\,,
\end{equation}
as well as four independent superspace coordinates $\eta^A$ satisfying
\begin{equation}
\{ \eta^A, \eta^B \} = 0\,.
\end{equation}
For more details on these conventions, see for instance~\cite{Elvang:2013cua}.

The conservation of momentum and half of the components of supermomentum (those corresponding to the generators $Q^{a A} = \lambda^a \eta^A$)\footnote{The conservation of the $\overline{Q}^{\dot{a}}_A = \widetilde{\lambda}^{\dot{a}} \frac{\partial}{\partial \eta^A}$ is not manifest in this formalism, but implies non-trivial and powerful differential constraints on superamplitudes~\cite{Caron-Huot:2011zgw}.} can be made manifest by writing the $n$-particle superamplitude ${\cal{A}}_n$ with an explicit prefactor of
\begin{equation}
\delta^4\left(\sum_{i=1}^n \lambda_i^a \widetilde{\lambda}_i^{\dot{a}}\right)
\delta^8\left(\sum_{i=1}^n \lambda_i^a \eta_i^A\right).
\label{eq:deltas}
\end{equation}
(The sole exception is the so-called $\overline{\rm MHV}$ three-point amplitude, which exists due to the peculiarities of massless three-point kinematics in four dimensions.) In light of Eq.~(\ref{eq:deltas}), the Grassmann Taylor expansion of ${\mathcal{A}}_n$ evidently begins at ${\mathcal{O}}(\eta^8)$, and, thanks to R-symmetry, can only contain terms with $8 + 4 k$ powers of $\eta$.  Terms in the expansion of ${\cal A}_n$ are conventionally denoted
\begin{equation}
\label{eq:pexpansion}
{\cal A}_n(\lambda_i^a, \widetilde{\lambda}_i^{\dot{a}}, \eta_i^A)
= \sum_{k = 0}^{n-4}
{\cal A}_n^{{\rm N}^{k}{\rm MHV}}(\lambda_i^a,
\widetilde{\lambda}_i^{\dot{a}}, \eta_i^A)\,,
\end{equation}
where ${\cal A}_n^{{\rm N}^{k}{\rm MHV}}$ is homogeneous of degree $4k+8$ in the $\eta$'s and MHV stands for maximally helicity violating.  Because of the overall supermomentum conserving delta function, we can say that ${\cal A}_n^{{\rm N}^{k}{\rm MHV}}$ is equal to $\delta^8(q)$ times a homogeneous polynomial in the $\eta$'s of degree $4k$.  The terms with $k=0,1,2,\ldots$ are referred to as MHV, NMHV (next-to-MHV), NNMHV (next-to-next-to-MHV), etc.

In the planar limit, where only single-trace color structures
contribute,\footnote{The color-stripped planar $n$-gluon amplitudes ${\cal A}_n$ are the coefficients of a single-trace color factor ${\rm Tr}(T^{a_1}T^{a_2}\cdots T^{a_n})$, and naturally transform under the dihedral group $D_n$.} it is useful to trivialize (super)momentum conservation by formulating this constraint geometrically.  If we place the $n$ four-momentum vectors $p_i^{a \dot{a}}$ of the scattering particles head to tail in the order dictated by the color trace, they form a closed polygon in Minkowski space with light-like edges.  Such a configuration may alternatively be described by the locations of its vertices, which we denote by $x_i$ and call dual coordinates.  Specifically we have
\begin{equation}
\label{eq:dualvar}
x^{a \dot{a}}_i - x^{a \dot{a}}_{i+1} = p^{a \dot{a}}_i\,,
\end{equation}
and we similarly have $n$ Grassmann dual coordinates $\theta_i^{a A}$ obeying
\begin{equation}
\theta_{i}^{aA} - \theta^{aA}_{i{+}1}
= \lambda_i^a \eta^A_i \qquad \text{(no sum on $i$).}
\end{equation}
This notation not only serves to trivialize (super)momentum conservation (via periodicity in $i \rightarrow i+n$),
\begin{equation}
\delta^4(p) \delta^8(q) =
\delta^4(x_{n+1} - x_1) \delta^8(\theta_{n+1} - \theta_1)\,,
\end{equation}
it also helps to expose a striking property of planar scattering amplitudes in SYM theory called dual (super)conformal symmetry~\cite{Drummond:2007au,Drummond:2008vq}---which is simply superconformal symmetry in $(x,\theta)$ space.

One of the most remarkable developments in the understanding of scattering amplitudes in SYM theory is the discovery of the amplitude/Wilson loop correspondence~\cite{Alday:2007hr,Drummond:2007aua,Brandhuber:2007yx,Drummond:2007cf,Drummond:2007au,Bern:2008ap,Drummond:2008aq,Alday:2008yw,Adamo:2011pv}.  We saw that it was natural to picture the kinematic configuration $(p_1,\ldots,p_n)$ of $n$ null momenta satisfying energy-momentum conservation as a polygon with light-like edges and vertices located at the $n$ dual coordinates $(x_1,\ldots,x_n)$.  Let $\langle W \rangle$ denote the expectation value of a Wilson loop associated to this polygon.  In its simplest form, the amplitude/Wilson loop correspondence is the statement of the exact equivalence
\begin{equation}
\label{eq:Wilsonloop}
\log \frac{{\cal{A}}_n^{\rm MHV}(p_1,\ldots,p_n)}{
{\cal{A}}_n^{\rm MHV}(p_1,\ldots,p_n)\rvert_{\rm tree-level}} =
\log\,\langle W(x_1,\ldots,x_n) \rangle
\end{equation}
in planar SYM theory.  A generalization of this formula is also known to hold for non-MHV amplitudes, when the Wilson loop is suitably decorated by the insertion of certain operators on its edges.

Both sides of Eq.~(\ref{eq:Wilsonloop}) are divergent.  The left-hand side has the usual infrared divergences of massless gauge theories, while the right-hand side has ultraviolet divergences arising from gluon exchange between adjacent edges of the polygon near its corners.  Fortunately the divergences of both sides are very well-understood and take a simple factorized form.  In dimensional regularization to $D = 4 - 2 \epsilon$ we can write~\cite{Magnea:1990zb,Catani:1998bh,Sterman:2002qn,Bern:2005iz}
\begin{equation}
\label{eq:logW}
\log\,\langle W(x_1,\ldots,x_n) \rangle =
\sum_{i=1}^n \Div(x_{i-1,i+1}^2; \epsilon)
+ \Fin_n(x_{ij}^2)
\end{equation}
where $x_{i,j}^2 = (x_i - x_j)^2$, 
\begin{equation}
\Div(x^2; \epsilon) = - \frac{1}{4} \sum_{L=1}^\infty
g^{2L} ( - x^2 \mu^2 )^{L \epsilon} \left[
\frac{\Gamma_{\rm cusp}^{(L)}}{(L \epsilon)^2} +
\frac{\Gamma_{\rm collinear}^{(L)}}{L \epsilon}
\right]
\end{equation}
and $\Fin_n(x_{ij}^2)$ is free of infrared and ultraviolet divergences.
For a gauge group $SU(N)$ and gauge coupling $g_{\rm YM}$, we define
\begin{equation}
g^2 \equiv \frac{g_{\rm YM}^2 N}{16 \pi^2} = \frac{\lambda}{16 \pi^2} \,,
\end{equation}
where $\lambda$ is the `t Hooft coupling of planar SYM theory, $\mu$ is an arbitrary mass parameter, and the two sequences of numbers denoted $\Gamma^{(L)}$ are respectively the $L$-loop cusp and collinear anomalous dimensions.\footnote{The collinear anomalous dimension for amplitudes differs from that for Wilson loops; the difference drops out for suitable finite ratios.}

The Wilson loop has conformal symmetry in $x$-space (this is the dual conformal symmetry of the corresponding amplitude), except that this is broken by the UV divergences at the cusps. In the $\epsilon \to 0$ limit, the breaking of the conformal symmetry manifests itself as an anomalous Ward identity~\cite{Drummond:2007au}
\begin{equation}
\label{eq:ward}
K^\mu \Fin_n(x_{ij}^2) = \frac{1}{2} \Gamma_{\text{cusp}}(g^2)
\sum_{i=1}^n x_{i,i+1}^\mu \log \left(  x_{i,i+2}^2/ x_{i-1,i+1}^2\right)
\end{equation}
for the generator of special conformal transformations
\begin{equation}
K^\mu = \sum_{i=1}^n \left[
2 x_i^\mu x_i^\nu \frac{\partial}{\partial x_{i \nu}} -
x_i^2 \frac{\partial}{\partial x_{i \mu}} \right].
\end{equation}
One particular solution to the conformal Ward identity~(\ref{eq:ward}) is a function of the $x_{ij}^2$ known as the BDS ansatz~\cite{Bern:2005iz}.\footnote{More precisely, the term ``BDS ansatz'' usually refers to the sum of this particular solution and the divergent terms displayed explicitly in Eq.~(\ref{eq:logW}).}  Therefore, the Ward identity completely determines the Wilson loop expectation value (and hence, the MHV amplitude) up to the addition of any homogeneous solution of Eq.~(\ref{eq:ward}), i.e.~up to any dual conformally invariant function of the $x_{ij}^2$.  The finite, dual conformally invariant quantity obtained by subtracting the BDS ansatz from Eq.~(\ref{eq:Wilsonloop}) is called the MHV remainder function $R_n$.  Because the polygon edges are light-like, $x_{i,i+1}^2=0$, it is impossible to form any non-trivial dual conformal cross ratios for $n<6$. So the first nontrivial instance of the remainder function is for $n=6$, where three independent cross ratios
\begin{equation} \label{eq:six_point_uvw}
u = \frac{x_{13}^2 x_{46}^2}{x_{14}^2 x_{36}^2}, \qquad
v = \frac{x_{24}^2 x_{51}^2}{x_{25}^2 x_{41}^2}, \qquad
w = \frac{x_{35}^2 x_{62}^2}{x_{36}^2 x_{52}^2}\,
\end{equation}
can be defined. In general, there are $3(n-5)$ independent variables,
and this is the dimensionality of the phase-space for $n$-point scattering
in planar SYM, five variables fewer than in a generic theory.

In practice, it is often convenient to parametrize dual-conformally-invariant kinematics using momentum twistors~\cite{Hodges:2009hk}, which are four-component objects defined by
\begin{equation}
Z^I_i = (\lambda_i^a, x_i^{b \dot{a} }\lambda_{i b})\, 
\end{equation}
for each particle index $i$, where $I= (a, \dot{a})$ is a combined $SU(2,2)$ index. Momentum twistors are invariant under overall rescalings $Z_i^R \to t_i Z_i^R$ and as such represent points in $\mathbb{CP}^3$. They also transform linearly under dual conformal transformations. In $n$-particle kinematics, they can be assembled into a $4 \times n$ matrix 
\begin{equation}
Z \in \text{Gr}(4,n)/ \text{GL}(1)^{n-1}\, ,
\end{equation}
which corresponds to a point in the Grassmannian of four-dimensional subspaces in $\mathbb{CP}^n$, modulo independent rescalings on its columns. Up to these rescalings, every element of $\text{Gr}(4,n)$ thus specifies a point in $n$-particle kinematics; in particular, the value of dual conformal cross ratios such as those shown in Eq.~\eqref{eq:six_point_uvw} can be computed by making the replacement
\begin{equation}
x_{ij}^2 \rightarrow \text{det} (Z_{i-1} Z_i Z_{j-1} Z_j ) \, ,
\end{equation}
as all other factors in these ratios cancel out. In the literature, these determinants are usually denoted by four-brackets as $\langle i j k l \rangle = \text{det} (Z_i Z_j Z_k Z_l)$.


\section{Amplitude Integrands and the Amplituhedron}
\label{sec:amplituheron}

At tree level, scattering amplitudes are rational functions of kinematical variables. In the planar limit they are especially simple, and poles can only appear when squared sums of consecutive momenta vanish, namely when $(p_i{+}p_{i{+}1}{+}{\cdots}{+}p_j)^2=0$. Amplitudes factorize on these poles into pairs of subamplitudes, as dictated by unitarity. At loop level, while amplitudes generally evaluate to complicated transcendental functions, it is possible (in the planar case) to define a rational $n$-point N$^k$MHV $\ell$-loop \emph{integrand} ${\cal I}_{n,k}^{\rm \ell-loop}$, which is a function of both external kinematics and loop momenta. We can think about these rational functions as the integrands one would get from summing over all Feynman diagrams prior to integration; however, because of the power-counting properties of $\mathcal{N}=4$ SYM theory, these functions are also uniquely determined by the requirement that they satisfy all possible cuts of the amplitude \cite{Arkani-Hamed:2010zjl}. In practice, this fact can be used to compute ${\cal I}_{n,k}^{\rm \ell-loop}$ much more efficiently than would be possible using Feynman diagrams. 

Although one must integrate over the loop momenta in these integrands to obtain the full loop-level amplitude, it also proves interesting to study these rational functions prior to integration. In doing so, it is usually advantageous to make use of momentum twistors, as defined in section~\ref{sec:background}. These variables make the dual conformal symmetry of planar ${\cal N}=4$ SYM amplitudes completely manifest, and furnish the space of kinematics with a nice geometric interpretation. Namely, in momentum twistor space, we have $n$ ordered momentum twistors $Z_i$ representing the external momenta, and $L$ lines $(AB)_j$ representing the independent loop momenta, which are each represented by a pair of points $Z_A$ and $Z_B$. In this framework, the cuts of the amplitude only have support on special configurations of the lines $(AB)_j$, in which they intersect the external lines $Z_iZ_{i{+}1}$ in a specified way. Given the function ${\cal I}_{n,k}^{\rm \ell-loop}$, which is constructed to match the predictions of field theory on all of the amplitude's cuts, the full integrand form $\Omega_{n,k}^{\rm \ell-loop}$ for the $n$-point N$^k$MHV amplitude at $\ell$ loops is given by 
\begin{equation}
\Omega_{n,k}^{\rm \ell-loop} = d\mu_1 d\mu_2\dots d\mu_\ell\,{\cal I}_{n,k}^{\rm \ell-loop} \, ,
\end{equation}
where $d\mu_k = \langle AB\,d^2A\rangle\langle AB\,d^2B\rangle$ for each loop momentum. In order to carry out these integrals over the loop momenta, one must generally regularize these integrals in the infrared. 

The integrand form $\Omega_{n,k}^{\rm \ell-loop}$ can also be obtained as the canonical differential form on the \emph{Amplituhedron geometry} \cite{Arkani-Hamed:2013jha,Arkani-Hamed:2013kca}. This geometry is defined as a special configuration of momentum twistors $Z_i$ and lines $(AB)_j$ which are subject to certain positivity conditions. For $\Omega_{n,k}^{\rm \ell-loop}$, the momentum twistors are chosen to satisfy
\begin{equation}
\langle i\,i{+}1\,j\,j{+}1\rangle>0 \, , \quad \mbox{where the series $\{\langle 1234\rangle, \langle 1235\rangle,\dots,\langle 123n\rangle\}$ has $k$ sign flips.}
\end{equation}
In addition, each loop momentum line $(AB)$ must satisfy  
\begin{equation}
\!\! \langle AB\,i\,i{+}1\rangle>0 \, , \quad \mbox{where the series $\{\langle AB12\rangle, \langle AB13\rangle,\dots,\langle AB1n\rangle\}$ has $k{+}2$ sign flips.}
\end{equation}
For each pair of lines $(AB)_i$, $(AB)_j$, we also require that $\langle (AB)_i(AB)_j\rangle>0$. This defines the loop Amplituhedron space ${\cal A}_{n,k}^{(\ell)}$. We can think about the Amplituhedron as being parametrized by a set of variables $x_i$ (the degrees of freedom in $Z_i$ and $(AB)_j$) which are subject to certain polynomial inequalities. The boundaries of the Amplituhedron correspond to the loci where either $\langle i\,i{+}1\,j\,j{+}1\rangle=0$ or $\langle (AB)_i\,j{+}1\rangle=0$, which are exactly the poles of the loop integrand ${\cal I}_{n,k}^{\rm \ell-loop}$. 

In general, we can define a canonical form $\smash{\omega_{n,k}^{\rm \ell-loop}}$ that has logarithmic singularities on the boundaries of the space $\smash{\cal A}_{n,k}^{(\ell)}$. Namely, this form has a singularity of the form $\smash{dx/x}$ whenever we approach one of the boundaries of the Amplituhedron corresponding to $x = 0$ (and nowhere else). This canonical form is guaranteed to exist for $\smash{\cal A}_{n,k}^{(\ell)}$, and to be unique. The amplitude form $\smash{\Omega_{n,k}^{\rm \ell-loop}}$ can then be obtained from $\smash{\omega_{n,k}^{\rm \ell-loop}}$ by a simple replacement
\begin{equation}
\Omega_{n,k}^{\rm \ell-loop} = \omega_{n,k}^{\rm \ell-loop}(dZ_i \rightarrow \eta_i) \, ,
\end{equation}
where the differentials $dZ_i$ are replaced by the fermionic variables $\eta_i$. This provides a completely geometric reformulation of ${\cal N}=4$ tree-level amplitudes, as well as the integrands of planar loop-level amplitudes. In particular, the normal physical properties of amplitudes, such as their singularity and branch cut structure, emerge as nontrivial consequences of the positivity conditions that define the Amplituhedron geometry. Moreover, the computational problem of obtaining a particular loop integrand $\Omega_{n,k}^{\rm \ell-loop}$ as a sum of Feynman diagrams, or as the product of recursion relations, is translated into the mathematical problem of triangulating the Amplituhedron space. One promising direction in the effort to extend the Amplituhedron picture to other theories is the formulation of the positive geometry in the Mandelstam or spinor helicity space \cite{Arkani-Hamed:2017mur,He:2018okq,Damgaard:2019ztj}.

While a closed-form expression for the integrand of planar amplitudes in $\mathcal{N}=4$ SYM theory remains an open problem, the geometric problem has been solved for some all-loop-order cuts \cite{Arkani-Hamed:2018rsk}. Moreover, the connection to Wilson loops and infrared-finite quantities has allowed for the development of interesting geometric approximations for all-loop quantities including the cusp anomalous dimension \cite{Arkani-Hamed:2021iya}. The Amplituhedron picture has also been studied extensively from the purely mathematical perspective \cite{Karp:2017ouj,Lukowski:2020dpn,Parisi:2021oql}, as it provides a substantial generalization of the positive Grassmannian \cite{Postnikov:2006kva,Arkani-Hamed:2012zlh}. As a mathematical structure, it is closely related to cluster algebras and other interesting algebraic structures that remain intact after one carries out the integration over loop momenta, as we discuss in the next section.


\section{Mathematical Properties at Loop Level}
\label{sec:mathematical_properties}

At loop level, amplitudes generally evaluate to transcendental functions that have an exceedingly complicated branch cut structure. Since our \emph{a priori} understanding of this analytic structure remains limited beyond special cases such as $2 \to 2$ scattering, most amplitudes of interest remain prohibitively difficult to evaluate using current technology. Despite this general situation, several infinite classes of amplitudes in planar $\mathcal{N}=4$ SYM theory have been uncovered over the last two decades whose analytic structure is simple enough to be understood either to all loop orders or at all particle multiplicity. In particular, the MHV amplitudes in this theory are known to two loops for any number of particles~\cite{Caron-Huot:2011zgw,Golden:2013xva,Golden:2014xqa}, while its six- and seven-particle amplitudes have been computed to high loop orders~\cite{Caron-Huot:2020bkp} and are not expected to exhibit new types of analytic structure at higher orders in perturbation theory.

One of the advantageous features of these classes of amplitudes is that they can be expressed in terms of multiple polylogarithms, or iterated integrals over logarithmic integration kernels~\cite{Chen,G91b,Goncharov:1998kja,Remiddi:1999ew,Borwein:1999js,Moch:2001zr}, whose properties as special functions are well understood. In particular, the analytic structure of multiple polylogarithms can be systematically exposed using the symbol~\cite{Goncharov:2010jf}, which maps multiple polylogarithms to a tensor product of logarithms that encodes all the logarithmic and algebraic branch of the original function.\footnote{Technically, this is only true modulo contributions proportional to transcendental constants such as $\zeta_3$; however, these terms can be captured as well by upgrading the symbol to a full coaction~\cite{Gonch2,Brown:2011ik,Brown1102.1312,Duhr:2012fh}.} Moreover, as the identities that hold between logarithms are completely understood (up to algebraic identities between their arguments), it is easy to find all identities between the symbols of multiple polylogarithms; these identities can then be uplifted to the original space of multiple polylogarithms through the inclusion of further contributions proportional to transcendental constants~\cite{Duhr:2011zq,Duhr:2012fh}.\footnote{We highlight that it can still prove highly nontrivial to find all identities between multiple polylogarithms when complicated algebraic functions appear in the arguments of these logarithms; see for instance~\cite{Bourjaily:2019igt}.} Algorithms also exist for systematically expressing multiple polylogarithms in terms of so-called fibration bases, which allow these functional relations to be imposed on an expression systematically~\cite{FBThesis,Anastasiou:2013srw,Panzer:2014caa}.  

A great deal has been learned about the specific logarithmic arguments---or symbol letters---that appear in polylogarithmic amplitudes in planar $\mathcal{N}=4$ sYM theory from the study of the theory's two-loop MHV amplitudes. In~\cite{Golden:2013xva} it was shown that the symbol letters that appear in the $n$-particle instance of this class of amplitudes can always be chosen to be cluster coordinates defined on the Grassmannian $\text{Gr}(4,n)$~\cite{1021.16017}. This initial observation has led to a prolific body of work that has tied the analytic structure of these amplitudes to cluster algebras~\cite{Golden:2014pua,Golden:2014xqa,Golden:2014xqf,Drummond:2017ssj,Bourjaily:2018aeq,Drummond:2018dfd,Golden:2018gtk,Drummond:2018caf,Golden:2019kks,Golden:2021ggj}, and to closely related algebraic structures such as tropical fans and polytopes~\cite{Drummond:2019qjk,Drummond:2019cxm,Arkani-Hamed:2019rds,Henke:2019hve,Drummond:2020kqg,Mago:2020kmp,Chicherin:2020umh,Mago:2020nuv,Herderschee:2021dez,He:2021esx,Mago:2021luw,Henke:2021ity,Ren:2021ztg,Yang:2022gko}. Of particular note are the cluster adjacency conditions~\cite{Drummond:2017ssj}, which state that symbol letters only appear in adjacent entries of the symbol of (appropriately-normalized) amplitudes when they also appear together in a cluster. In all known cases, this requirement has been observed to be equivalent to the implications of the extended Steinmann relations~\cite{Steinmann,Steinmann2,Cahill:1973qp,Caron-Huot:2016owq,Caron-Huot:2019bsq}, which restrict these amplitudes from having nonzero double discontinuities in partially-overlapping momentum channels (at any depth in the symbol). Similar constraints on the analytic properties of amplitudes have also been deduced more directly from the Landau equations~\cite{landau1959} and cut integrals~\cite{Cutkosky:1960sp}; see for instance~\cite{Abreu:2014cla,Dennen:2015bet,Bloch:2015efx,Kreimer:2016tqq,Dennen:2016mdk,Prlina:2017azl,Abreu:2017ptx,Prlina:2017tvx,Prlina:2018ukf,Bourjaily:2020wvq,Benincasa:2020aoj,Hannesdottir:2021kpd,Hannesdottir:2022bmo} for recent work in this direction.

The six- and seven-particle amplitudes in planar $\mathcal{N}=4$ SYM theory are especially simple, insofar as the cluster coordinates defined on $\text{Gr}(4,6)$ and $\text{Gr}(4,7)$ appear to describe the full set of symbol letters that appear in these amplitudes at any loop order. As we will review in section~\ref{sec:bootstrap}, this expectation has been leveraged to bootstrap the six-particle amplitude through seven loops~\cite{Dixon:2011pw,Dixon:2011nj,Dixon:2013eka,Dixon:2014voa,Dixon:2014iba,Dixon:2015iva,Caron-Huot:2016owq,Dixon:2016apl,Caron-Huot:2019vjl,Dixon:2020cnr}, and the seven-particle through four loops~\cite{Drummond:2014ffa,Dixon:2016nkn,Drummond:2018caf,Dixon:2020cnr}. Access to such high-loop data has, in turn, made it possible to uncover additional types of structure in these amplitudes, such as interesting number-theoretic symmetries under the cosmic Galois group~\cite{Cartier2001,2008arXiv0805.2568A,2008arXiv0805.2569A,Brown:2015fyf}. These symmetries restrict what numerical constants are expected appear in these amplitudes perturbatively; for instance, the constant $\zeta_3$ is not expected to appear in the six-particle amplitude at a particular kinematic point, at any loop order, when it is `cosmically normalized' in the way described in~\cite{Caron-Huot:2019bsq}. Notably, similar number-theoretic symmetry properties have been observed in massless $\phi^4$ theory~\cite{Schnetz:2013hqa,Panzer:2016snt}, QED~\cite{Laporta:2017okg,Schnetz:2017bko}, and string theory~\cite{Schlotterer:2012ny}.

Functions beyond multiple polylogarithms also appear in planar $\mathcal{N}=4$ SYM theory, for instance at two loops for ten particles~\cite{Caron-Huot:2012awx,Paulos:2012nu,Nandan:2013ip,Bourjaily:2017bsb,Kristensson:2021ani,Wilhelm:2022wow}. A great deal of work has gone into understanding the next class of special functions that naturally arises, which involves integrals over elliptic curves~\cite{SABRY1962401,Broadhurst:1993mw,Berends:1993ee,Bauberger:1994by,Bauberger:1994hx,Bauberger:1994nk,Laporta:2004rb,Groote:2005ay,MullerStach:2012az,Groote:2012pa,Bloch:2013tra,Adams:2013kgc,Adams:2014vja,Adams:2015gva,Adams:2015ydq,Adams:2016xah,Remiddi:2016gno,Adams:2017tga,Adams:2017ejb,Bogner:2017vim,Remiddi:2017har,Bourjaily:2017bsb,Broedel:2017siw,Chen:2017soz,Adams:2018yfj,Broedel:2018iwv,Adams:2018bsn,Adams:2018kez,Honemann:2018mrb,Bogner:2019lfa,Broedel:2019hyg,Broedel:2019kmn,Bourjaily:2020hjv,Bourjaily:2021vyj,Kristensson:2021ani,Wilhelm:2022wow}. Even more complicated integrals, such as integrals over hyperelliptic curves~\cite{Huang:2013kh,Hauenstein:2014mda} and over Calabi-Yau manifolds of unbounded dimension~\cite{Groote:2005ay,Brown:2010bw,Bloch:2014qca,Bloch:2016izu,mirrors_and_sunsets,Primo:2017ipr,Bourjaily:2018ycu,Bourjaily:2018yfy,Bourjaily:2019hmc,Klemm:2019dbm,Bonisch:2020qmm,Bonisch:2021yfw} also appear. The evaluation of amplitudes that involve functions more complicated than multiple polylogarithms remains an active area of research; see~\cite{Bourjaily:2022bwx} for a white paper devoted to this topic, and~\cite{Weinzierl:2022eaz} for a more pedagogical introduction.


\section{Special Kinematic Limits}
\label{sec:kinematic_limits}

Much more is known about the structure of amplitudes in $\mathcal{N}=4$ SYM theory in special kinematic limits. One important class of examples are multi-Regge limits, where all outgoing particles are strongly ordered in rapidity. Amplitudes exponentiate in these limits and admit an effective description as an expansion in large logarithms.  This exponentiation is especially well understood in the planar limit of this theory~\cite{Bartels:2009vkz,Fadin:2011we,Bartels:2011ge,Lipatov:2012gk,Dixon:2012yy,Bartels:2013jna,Basso:2014pla,Dixon:2014iba,DelDuca:2016lad,DelDuca:2018hrv,DelDuca:2019tur}, where it has been shown that the coefficients multiplying these large logarithms are always expressible in terms of specific classes of single-valued multiple polylogarithms~\cite{Dixon:2012yy,DelDuca:2016lad}. Moreover, predictions for these expansions are available at all loop orders and for any number of particles~\cite{Basso:2014pla,DelDuca:2019tur}. For a recent introduction to this topic, see for instance~\cite{DelDuca:2022skz}. 

Another interesting limit that has been studied in great detail is the near-collinear limit of planar amplitudes in $\mathcal{N}=4$ SYM theory. In the dual theory, this limit admits a non-perturbative description in terms of the so-called pentagon operator product expansion (POPE)~\cite{Alday:2010ku,Basso:2013vsa,Basso:2013aha,Basso:2014koa,Basso:2014jfa,Basso:2014nra,Belitsky:2014sla,Belitsky:2014lta,Basso:2014hfa,Belitsky:2015efa,Basso:2015rta,Basso:2015uxa,Belitsky:2016vyq}, by means of which the amplitude can be computed as an expansion in terms of flux-tube excitations crossing the Wilson loop. While in principle the POPE encodes the full amplitude, it is not yet known how to resum this expansion beyond one loop~\cite{Lam:2016rel,Bork:2019aud}. A form factor operator product expansion (FFOPE) has recently been developed in planar $\mathcal{N}=4$ SYM theory~\cite{Sever:2020jjx,Sever:2021nsq,Sever:2021xga}, which leverages the duality between form factors and Wilson loops in a periodic target space~\cite{Alday:2007he,Maldacena:2010kp,Brandhuber:2010ad}.

Further limits have also been studied in six- and seven-particle kinematics, where a number of amplitudes have been computed in general kinematics. These include multi-particle factorization limits~\cite{Dixon:2015iva,Dixon:2016nkn}, and the origin of the six-particle amplitude, which is conjecturally known to all loop orders~\cite{Basso:2020xts}. The amplitude also becomes singular when the dual Wilson polygon crosses itself, and an evolution equation has been derived that governs these singularities, as well as a proposed all-orders resummation~\cite{Dixon:2016epj,Caron-Huot:2019vjl}.


\section{Perturbative Bootstrap Calculations}
\label{sec:bootstrap}

The BDS ansatz for planar amplitudes in $\mathcal{N}=4$ SYM theory needs to be corrected for amplitudes involving more than five particles, starting at two loops. As reviewed in section~\ref{sec:background}, these corrections take the form of finite functions of dual-conformally-invariant cross ratios. In an impressive calculation, the first nontrivial correction---the two-loop correction to the six-particle amplitude---was integrated directly using  Mellin-Barnes techniques, and found to be expressible in terms of multiple polylogarithms~\cite{DelDuca:2009au,DelDuca:2010zg}. Symbol methods were subsequently used to put this function into a more parsimonious form, which makes it clear that only nine symbol letters appear~\cite{Goncharov:2010jf}:
\begin{equation}
\mathcal{S}_6 = \{u, v, w, 1-u, 1-v, 1-w, y_u, y_v, y_w \} \, ,
\end{equation}
where $u$, $v$, and $w$ were defined in~\eqref{eq:six_point_uvw}, and 
\begin{equation}
y_u = \frac{u-z_+}{u-z_-}\,, \qquad y_v = \frac{v-z_+}{v-z_-}\,,
\qquad y_w = \frac{w - z_+}{w - z_-}\, ,
\end{equation}
where
\begin{equation}
z_\pm = \frac{1}{2}\Bigl[-1+u+v+w \pm \sqrt{\Delta}\Bigr],
\qquad
\Delta = (1-u-v-w)^2 - 4 u v w \, .
\end{equation}
In other words, this function only develops logarithmic branch cuts on the nine codimension-one surfaces where these letters vanish in the space of dual-conformally-invariant cross ratios.

Equipped with this insight into the analytic structure of the two loop contribution, a bootstrap approach to computing the six-particle remainder function at higher loops was initiated in~\cite{Dixon:2011pw}. This approach starts from the assumption that no further symbol letters appear in the dual-conformally-invariant correction to the BDS ansatz, and tries to identify the unique polylogarithmic function that has all the right properties to encode the amplitude. This assumption turns out to be valid, and bootstrap methods have now been used to determine the amplitude through seven loops~\cite{Dixon:2011pw,Dixon:2011nj,Dixon:2013eka,Dixon:2014voa,Dixon:2014iba,Dixon:2015iva,Caron-Huot:2016owq,Dixon:2016apl,Caron-Huot:2019vjl}. As part of this work, the mathematical properties of the six-particle amplitude have been studied in great depth, and are now known to include:
\begin{enumerate}
\item[(i)] {\bf Dihedral Symmetry --} The amplitude is invariant under relabelings of its external legs that respect the original planar ordering.
\item[(ii)] {\bf Branch Cut Conditions --} When formulated in the Euclidean region, the amplitude should only develop branch cuts where one of the Mandelstam invariants vanishes or approaches infinity.
\item[(iii)] {\bf Final Entry Conditions --} Only certain letters are allowed to appear in the last entry of the symbol, as prescribed by the action of the $\bar{Q}$ equation~\cite{Caron-Huot:2011zgw,Caron-Huot:2011dec,Bullimore:2011kg}. 
\item[(iv)] {\bf Extended Steinmann Relations --} When appropriately normalized, the amplitude never involves sequential discontinuities in partially-overlapping three-particle momentum channels~\cite{Steinmann,Steinmann2,Cahill:1973qp,Caron-Huot:2019bsq}. This turns out to be equivalent to the cluster adjacency conditions proposed in~\cite{Drummond:2017ssj}.
\item[(v)] {\bf Cosmic Galois Coaction Principle --} The span of functions of fixed weight that appear in the first entry of the coaction of the amplitude stabilizes after a certain number of loop orders; this observation restricts the space of functions that are expected to show up to all higher loop orders~\cite{Caron-Huot:2019bsq}. 
\item[(vi)] {\bf Multi-Regge Kinematics --} In this limit, the outgoing particles are strongly ordered in rapidity and the amplitude exponentiates. It can be independently computed as an expansion in large logarithms using an effective description in terms of an impact factor and BFKL eigenvalue~\cite{Bartels:2009vkz,Fadin:2011we,Bartels:2011ge,Lipatov:2012gk,Dixon:2012yy,Basso:2014pla,Dixon:2014iba}.   
\item[(vii)] {\bf Near-Collinear Kinematics --} The expansion of the amplitude around collinear limits is described at finite coupling by the POPE~\cite{Alday:2010ku,Basso:2013vsa,Basso:2013aha,Basso:2014koa,Basso:2014jfa,Basso:2014nra,Belitsky:2014sla,Belitsky:2014lta,Basso:2014hfa,Belitsky:2015efa,Basso:2015rta,Basso:2015uxa,Belitsky:2016vyq}, which makes it possible to independently compute the first terms in this near-collinear expansion at fixed loop order.
\item[(viii)] {\bf Self-Crossing Kinematics --} Kinematics in which the transverse momentum of a pair of outgoing gluons vanishes and the amplitude becomes singular. The singular terms are governed by an evolution equation, and have been determined to high loop order~\cite{Dixon:2016epj}.
\item[(ix)] {\bf Behavior Near the `Origin' --}  The behavior of the MHV amplitude near the `origin' of six-particle kinematics, where $u$, $v$, and $w$ all vanish, is conjecturally understood to all loop orders~\cite{Basso:2020xts}, and it is the exponential of a quadratic form in the logarithms of $u,v,w$. 
\end{enumerate}
It is believed that the six-particle amplitude is determined by (a subset of) these constraints to all orders in perturbation theory; however, in practice constructing the explicit functions becomes too computationally intensive beyond seven or eight loops. To illustrate the power of this procedure, we present the number of free parameters that remain at various steps in the bootstrap procedure for the MHV amplitude through six loops in Table~\ref{tab:six_MHV}. After the amplitude has been uniquely determined, the remaining constraints act as cross checks.  The NMHV helicity configuration works in a similar fashion.

\renewcommand{\arraystretch}{1.25}
\begin{table}[!t]
\centering
\begin{tabular}[t]{l c  c c  c c c c c}
\hline
Constraint                      & $L=1$\, & $L=2$\, & $L=3$\, & $L=4$\, & $L=5$\, & $L=6$
\\\hline
1. $\Hhex$ & 6 & 27 & 105 & 372 & 1214 & 3692?\\\hline\hline
2. Symmetry       & 2 & 7 & 22 &  66 & 197 & 567 \\\hline
3. Final-entry       & 1 & 4 & 11 &  30 & 85 & 236 \\\hline
4. Collinear       & 0 & 0 & $0^*$ &  $0^*$ & $1^{*3}$ & $6^{*2}$ \\\hline
5. LL MRK       & 0 & 0 & 0 &  0 & $0^*$ & $1^{*2}$ \\\hline
6. NLL MRK       & 0 & 0 & 0 &  0 & $0^*$ & $1^*$ \\\hline
7. NNLL MRK       & 0 & 0 & 0 &  0 & 0 & $1$ \\\hline
8. N$^3$LL MRK       & 0 & 0 & 0 &  0 & 0 & 1 \\\hline
9. Full MRK       & 0 & 0 & 0 &  0 & 0 & 1 \\\hline
10. $T^1$ OPE       & 0 & 0 & 0 &  0 & 0 & 1 \\\hline
11. $T^2$ OPE       & 0 & 0 & 0 &  0 & 0 & 0 \\\hline
\end{tabular}
\caption{The number of free parameters that remain in the BDS-like normalized ans\"{a}tze for
the MHV six-particle amplitude after each constraint is applied. The initial ansatz is formed out of a general linear combination of the functions in the $\Hhex$ space, which includes all polylogarithms that involve just the letters in $\mathcal{S}_6$, and that satisfy conditions (ii), (iv), and (v). The superscripts ``$*$'' (or ``$*n$'') denote an additional
ambiguity (or $n$ ambiguities) that arises due to further ambiguities in the cosmic normalization constant $\rho$. The ``$?$'' indicates an ambiguity about
the number of weight 12 odd functions that are ``dropouts'', namely that are allowed at symbol level but not function level. The numbers in this table were taken from~\cite{Caron-Huot:2019vjl}, where further details can be found.}
\label{tab:six_MHV}
\end{table}

A similar bootstrap approach has also been employed to compute the seven-particle amplitude and certain three-point form factors in planar SYM. In the former case, 42 symbol letters appear in the two loop MHV amplitude~\cite{Caron-Huot:2011zgw}, and the same letters have been found to be sufficient for expressing both the MHV and NMHV amplitudes through four loops~\cite{Drummond:2014ffa,Dixon:2016nkn,Drummond:2018caf,Dixon:2020cnr}. In the latter case only six symbol letters appear, and the form factor has been bootstrapped through eight loops~\cite{Brandhuber:2012vm,Dixon:2020bbt,Dixon:2022rse}. Surprisingly, this form factor has also been shown to be dual to the six-particle amplitude evaluated on a two-dimensional kinematic surface, order-by-order in perturbation theory~\cite{Dixon:2021tdw}.  The duality reverses all entries in the symbol (the antipode map). It is not known yet whether antipodal duality appears in a wider class of processes.

More generally, while the planar two-loop $n$-particle MHV amplitude is known to involve $\frac{3}{2} n^3- 15 n^2 + \frac{77}{2} n$ letters~\cite{Caron-Huot:2011zgw}, additional letters are expected to appear at higher loops for eight or more particles. Such letters explicitly appear in the three-loop eight-point MHV amplitude, which was recently computed with the help of the $\bar{Q}$ equation~\cite{Li:2021bwg}. This fact makes bootstrap computations harder to pursue for more than seven particles, since there doesn't yet exist a reliable method for predicting the symbol letters that will appear in these amplitudes (although much work has been devoted to this question; see for instance~\cite{Arkani-Hamed:2012zlh,Golden:2013xva,Golden:2014xqa,Golden:2014pua,Drummond:2017ssj,Drummond:2018dfd,Golden:2018gtk,Drummond:2018caf,Golden:2019kks,Drummond:2019qjk,Drummond:2019cxm,Arkani-Hamed:2019rds,Henke:2019hve,Drummond:2020kqg,Mago:2020kmp,Chicherin:2020umh,Mago:2020nuv,Herderschee:2021dez,He:2021esx,Mago:2021luw,Henke:2021ity,Ren:2021ztg}). Further data on this question can be gathered by computing three-loop MHV amplitudes at higher points, which should also be possible with the help of the $\bar{Q}$ equation, using input from our knowledge of this theory's two-loop NMHV amplitudes~\cite{He:2019jee,He:2020vob}.

At higher points, amplitudes and form factors in $\mathcal{N}=4$ SYM theory are expected to involve functions beyond multiple polylogarithms, even in the planar limit. While it is expected that bootstrap approaches can also be applied to amplitudes that involve these more general types of functions---as indeed, these quantities are expected to exhibit many of the same algebraic and analytic features as the amplitudes that have already been bootstrapped---more technology for dealing with these functions is needed to make this approach feasible. Notably, however, a great deal of progress in this direction has recently been made in the case of elliptic polylogarithms, and similar advances are expected in the coming years in our understanding of the more general types of integrals that appear~\cite{Bourjaily:2022bwx}.


\section{Outlook}
\label{sec:outlook}

While an impressive amount is already known about the amplitudes in $\mathcal{N}=4$ SYM theory, there are many directions in which our understanding can be improved. We now highlight some of the important 
questions and research directions in which we expect progress can be made in the coming years:
\begin{itemize}
\item Much of the recent progress in this theory has been made on the planar limit, where significant simplifications occur. While initial results have also been achieved for four- and five-particle amplitudes with full color dependence~\cite{Henn:2016jdu,Abreu:2018aqd,Chicherin:2018yne}, it will become increasingly important to develop tools that scale well with the number of kinematic variables in order to compute amplitudes at higher points.
  
\item There is an ongoing search for the putative dual Amplituhedron, whose volume (rather than associated differential form) should reproduce tree-level amplitudes and loop integrands. The dual geometry for NMHV tree-level amplitudes was discovered some time ago~\cite{Hodges:2009hk,Arkani-Hamed:2010wgm}, and some encouraging results are available in the literature~\cite{Arkani-Hamed:2014dca,Ferro:2015grk,Herrmann:2020qlt,Herrmann:2020oud}; however, an explicit and general construction is still missing.

\item Another important direction is to extend the positive geometry construction beyond the planar limit. The Grassmannian formulation for on-shell diagrams \cite{Arkani-Hamed:2012zlh} does extend to non-planar diagrams \cite{Arkani-Hamed:2014bca,Franco:2015rma,Bourjaily:2016mnp,Herrmann:2016qea,Heslop:2016plj}, and there is evidence that many of the analytic properties that have been observed in the planar limit also persist outside of this limit, such as the absence of poles at infinity and the existence of only logarithmic singularities in the integrand~\cite{Arkani-Hamed:2014via,Bern:2014kca,Bern:2015ple,Bourjaily:2019iqr,Bourjaily:2019gqu}. However, how to uniquely define the non-planar integrand, and whether it can be geometrically formulated, remain important open questions.

\item As the POPE gives a non-perturbative formulation of this theory's planar amplitudes as expansions near two-particle collinear limits, it should in principle be possible to resum the contributions at fixed loop order. Currently the technology for doing this only exists at one loop~\cite{Lam:2016rel,Belitsky:2017wdo,Bork:2019aud,Bork:2020aut}. Barring a full resummation algorithm, it would also be interesting to be able to read properties of the amplitudes off of these sums, such as what combinations of kinematic variables these amplitudes depend on.
\item While the general class of special functions that can appear in perturbative quantum field theory remains unclear, this question can be answered for specific classes of amplitudes with the help of integrand-level basis reduction techniques~\cite{Melrose:1965kb,Passarino:1978jh,Bern:1994zx,Bern:1994cg,Britto:2004nc,Ossola:2006us,Mastrolia:2010nb,Bourjaily:2013mma,Bourjaily:2015jna,Bourjaily:2017wjl,Bourjaily:2020qca}. It would be interesting to catalog what types of functions might appear in $\mathcal{N}=4$ SYM theory at a given multiplicity or loop order.
\item The bootstrap approach described in section~\ref{sec:bootstrap} has proven to be wildly successful, and has given rise to some of the highest-order results with nontrivial kinematic dependence in any quantum field theory. However, these techniques currently remain restricted to functions that can be expressed in terms of multiple polylogarithms. As such, it will be important to extend them to function spaces involving elliptic curves and other higher-dimensional varieties; see the Snowmass white paper~\cite{Bourjaily:2022bwx}. A preliminary step could be to bootstrap higher-point amplitudes first on suitable lower-dimensional surfaces.
\item One of the main long-term goals in understanding the amplitudes of $\mathcal{N}=4$ SYM theory is to find a closed-form representation of even the simplest amplitudes at finite coupling. (One could argue this has already been done, up to constants, for the four- and five-particle amplitude in the form of the BDS ansatz~\cite{Bern:2005iz}, but these amplitudes benefit an exceptional amount from dual conformal symmetry~\cite{Drummond:2006rz, Bern:2006ew, Bern:2007ct, Alday:2007hr, Bern:2008ap, Drummond:2008vq}.) Some hints as to what form amplitudes might take at finite coupling come from resumming ladder integrals~\cite{Caron-Huot:2018dsv}; it would already be extremely interesting to find a similar finite-coupling formulation of the planar six-particle amplitude, or the planar three-particle form factor studied in~\cite{Brandhuber:2012vm,Dixon:2020bbt,Dixon:2022rse}. Such a formulation would represent a resummation of the POPE mentioned above.
\item Amplitudes and form factors have been observed to exhibit interesting number-theoretic properties under the Galois coaction, not only in $\mathcal{N}=4$ SYM theory~\cite{Caron-Huot:2019bsq} but also in massless $\phi^4$ theory~\cite{Schnetz:2013hqa,Panzer:2016snt}, electromagnetism~\cite{Laporta:2017okg,Schnetz:2017bko}, and string theory~\cite{Schlotterer:2012ny}.  It would be useful to find a physical explanation for these number-theoretic properties, so as to be able to make predictions about the number-theoretic properties of as-yet-uncomputed amplitudes. 
\item The $\bar{Q}$ equation has been utilized in a number of impressive calculations to compute amplitudes at all multiplicity~\cite{Caron-Huot:2011zgw,Caron-Huot:2011dec,He:2020vob,Li:2021bwg}. However, its utility currently remains limited to the MHV and NMHV sectors, as amplitudes in other sectors involve contributions that are in the kernel of the $\bar{Q}$ equation. It would be extremely interesting to understand these additional contributions in order to extend the reach of these methods.
\item The analytic structure of amplitudes is not well understood beyond four-particle scattering. One step that would improve our understanding of their analytic structure in the case of $\mathcal{N}=4$ SYM theory would be to elucidate the physics behind the connection between the singularity structure of some of its amplitudes and cluster algebras (and geometric structures closely related to cluster algebras, such as tropical fans and polytopes)~\cite{Arkani-Hamed:2012zlh,Golden:2013xva,Golden:2014xqa,Golden:2014pua,Drummond:2017ssj,Drummond:2018dfd,Golden:2018gtk,Drummond:2018caf,Golden:2019kks,Drummond:2019qjk,Drummond:2019cxm,Arkani-Hamed:2019rds,Henke:2019hve,Drummond:2020kqg,Mago:2020kmp,Chicherin:2020umh,Mago:2020nuv,Herderschee:2021dez,He:2021esx,Mago:2021luw,Henke:2021ity,Ren:2021ztg}. Progress is also being made on how the analytic structure of generic Feynman integrals can be better understood using the Landau equations~\cite{landau1959} and cut integrals~\cite{Cutkosky:1960sp}; see for instance~\cite{Abreu:2014cla,Dennen:2015bet,Bloch:2015efx,Kreimer:2016tqq,Dennen:2016mdk,Prlina:2017azl,Abreu:2017ptx,Prlina:2017tvx,Prlina:2018ukf,Bourjaily:2020wvq,Benincasa:2020aoj,Hannesdottir:2021kpd,Hannesdottir:2022bmo}. 
\item Almost all of the work on amplitudes in $\mathcal{N}=4$ SYM theory has been carried out at the origin of the theory's moduli space (although see~\cite{Caron-Huot:2014gia,Sakata:2017pue,Herderschee:2019dmc}). It would thus be interesting to better understand how the mathematical structure of its amplitudes change when some of its scalars are given a vacuum expectation value.
\end{itemize}
Of course, this list highlights just some of the topics that merit study over the next decade. In particular, we expect that many unanticipated research directions will arise with the identification of further types of mathematical structure in $\mathcal{N}=4$ SYM theory. With luck, some of these discoveries will give us a glimpse into the appropriate mathematical language in which our current 
descriptions of amplitudes in this theory can be seen to be encoded in expressions that are valid both at finite coupling and in general kinematics.

\section*{Acknowledgements}

We thank Zvi Bern and Benjamin Basso for stimulating discussions. This work is supported in part by the US Department of Energy under contracts DE--SC0009988 (N.A-H.), DE--AC02--76SF00515 (L.D.), DE--SC0010010 (M.S.~and A.V.), DE--SC0009999 (J.T.), in part by the funds of the University of California (J.T.), and in part by Simons Investigator Award \#376208 (A.V.).


\bibliographystyle{JHEP}
\bibliography{neq4_white_paper}

\end{fmffile}
\end{document}